\documentclass[10pt,a4paper]{article}
\usepackage{RR}
\usepackage{hyperref}

\usepackage{graphicx,psfig,amsmath,url,algorithmic,float,bbm}
\usepackage{times}
\usepackage{fancybox}
\usepackage[english]{babel}

\newcommand{\mybox}[2]{\begin{center}\boxed{\begin{array}{p{#1}}{#2}\end{array}}\end{center}}

\def\junk#1{}

\RRdate{August 2006}
\RRauthor{Dinesh Kumar 
  \and 
Dhiman Barman
  \and 
Eitan Altman
  \and 
Jean-Marc Kelif
}
\authorhead{Dinesh Kumar, Dhiman Barman, Eitan Altman \& Jean-Marc Kelif}
\RRtitle{Nouvelle Politique Inter-Couche de Commutation pour Transmission TCP sur 3G UMTS Downlink}
\RRetitle{New Cross-Layer Channel Switching Policy for TCP Transmission on 3G UMTS Downlink}
\titlehead{New Cross-Layer Channel Switching Policy for TCP Transmission on 3G UMTS Downlink}
\RRnote{Dinesh Kumar and Eitan Altman are at INRIA, BP 93, 06902 Sophia Antipolis, France.
\\
(\{dkumar, altman\}@sophia.inria.fr)}
\RRnote{Jean-Marc Kelif is at France Telecom R\&D, 92794 Issy les Moulineaux, France.
\\
(jeanmarc.kelif@orange-ft.com)}
\RRnote{Dhiman Barman, who is now at Computer Sc. \& Engg., University of California, Riverside, USA,
carried out his part of this work during a visit to INRIA. (dhiman@cs.ucr.edu)}

\RRnote{The authors of this report are thankful to France Telecom R\&D who financially
supported this work under the external research contract (CRE) no. 46130622.}

\RRresume{En 3G UMTS, deux canaux principaux de transport ont été crées pour la 
transmission de données sur le lien descendant : un canal commun FACH et un canal dédié DCH. 
Le performance de TCP en UMTS dépend beaucoup de la politique de commutation de canal 
utilisée. Dans cet article, nous proposons et analysons trois nouvelles politiques de 
commutations de canal pour l'UMTS basées sur des seuils simples que nous appelons politique 
QS (Queue Size), FS (Flow Size) et QSFS (QS et FS combinés). Ces politiques améliorent de 
manière significative (environ $17\%$) le temps de réponse par rapport à la politique 
modifiée de seuil de \cite{prabhu}. De plus, nous proposons et évaluons une nouvelle politique de 
commutation, plus performante, que nous appelons politique de FS-DCH ({\em at-least} flow-size threshold 
on DCH). Cette politique favorise les connexions TCP courtes (i.e., de peu de 
paquets) et est ainsi une politique {\em inter-couche} qui améliore le performance de TCP 
en donnant priorité aux premiers paquets des connexions sur le canal rapide DCH. 
De nombreux résultats de simulation illustrent ce gain de performance dans le cas 
d'un petit nombre de connexions TCP.}

\RRabstract{In 3G UMTS, two main transport channels have been provided for downlink
data transmission: a common FACH channel and a dedicated DCH channel.
The performance of TCP in UMTS depends much on the channel switching
policy used. In this paper, we propose and analyze three new basic
threshold-based channel switching policies for UMTS that we name as
QS (Queue Size), FS (Flow Size) and QSFS (QS \& FS combined) policy.
These policies significantly improve over a
modified threshold policy in \cite{prabhu} by about $17\%$ in response time metrics.
We further propose and evaluate a new improved switching
policy that we call FS-DCH ({\em at-least} flow-size threshold on
DCH) policy. This policy is biased towards short TCP flows\footnote{A {\em flow}
is defined as a burst of packets in a TCP connection.} of few packets
and is thus a {\em cross-layer} policy that improves the performance of TCP
by giving priority to the initial few packets of a flow on the fast DCH channel. 
Extensive simulation results confirm this improvement for the case 
when number of TCP connections is low.}
\RRmotcle{inter-couche, ordonnancement, ressource, gestion de buffer, DCH, FACH}
\RRkeyword{cross-layer, scheduling, resource, buffer management, DCH, FACH}
\RRprojets{Maestro}
\RRtheme{\THCom} 
\URSophia 
\begin{document}
\makeRR   
\section{Introduction}

Keeping in pace with the increasing demand from users for access to
information and services on public and private networks, the third
generation (3G) Universal Mobile Telecommunication System (UMTS) 
has been designed to offer services such
as high speed Internet access, high quality image and video exchange
and global roaming. Data traffic in UMTS has been classified broadly
into four different classes, namely--conversational, streaming,
interactive (e.g., web browsing) and background (e.g., email)
classes. The bulk of data in streaming and interactive transmissions
is carried over the {\em downlink} from UTRAN (UMTS Terrestrial
Radio Access Network) to a UE (User Equipment). Data generated in
the higher layers of UTRAN is carried over the air interface to the
UEs via the downlink transport channels, which are
mapped in the physical layer to different physical channels. There
are two types of layer-2 downlink transport channels that have been provided
in UMTS: {\it dedicated} channels and {\it common} channels. A common channel is a
resource shared between all or a group of users in a cell, where as a
dedicated channel is a resource identified by a certain code on a
certain frequency and is reserved for a single user only. The only
dedicated channel is termed as DCH and one of the six common
transport channels that is mainly used for packet data on the
downlink is the FACH channel \cite{WCDMA}. The number of DCH channels in a UMTS
cell is {\it interference limited}. If a new user's
connection cannot be admitted into the cell (this is decided by an
appropriate interference based CAC or connection admission control),
it must wait until a DCH channel is released by the already connected
users or until when interference conditions become suitable for this
new user to be allocated a new DCH channel. Being a dedicated
channel, DCH guarantees higher data rates but the set-up time for
DCH is significant (of the order of $250$ms). On the other hand, the
common channel FACH inherently guarantees lower data rates but its
set-up time is less. According to the WCDMA (Wideband-CDMA)
specifications detailed by the 3GPP group, for a particular user,
long flows with large amount of packets can be transmitted on the
user dedicated DCH channel and short flows of few packets can be
transmitted on the common FACH channel which is shared by all users.
However, the 3GPP specifications do not provide any standardization
of such a channel selection/switching policy. A network operator is
free to choose its own proprietary channel switching policy.

\subsection{Main Contributions}
\label{main-contributions}
In this paper, we propose some new {\em basic} channel switching policies
and analyze their performance characteristics through simulations. These new
policies are based on an extension of the modified threshold policy in \cite{prabhu}.
In the simulation results (Section \ref{evaluation}) we observe that our new switching policies improve
on the modified threshold policy in \cite{prabhu} by around $17\%$
in response time metrics. Based on some observations about the DCH
and FACH channel characteristics and the need for distinction of
long and short TCP flows, we further propose another new {\em cross-layer} channel
switching policy, which is our main contribution in this paper. 
To the best of our knowledge, ours is the first attempt to propose 
such a cross-layer channel switching policy for UMTS downlink that is 
based on diffrentiation between long and short TCP flows.
All the new policies are in accordance with the current WCDMA
specifications and we evaluate their performance in terms of
response time and slowdown using simulations.

\subsection{Synopsis}
\label{synopsis}
We start in Section~\ref{policy} by defining the basic
threshold-based channel switching policies. We name them as QS
(Queue Size) policy, FS (Flow Size) policy and QSFS (QS \& FS
combined) policy. In all these policies a new flow or connection
starts on the common FACH channel. In Section~\ref{newpolicy}, we
observe that in the basic policies the switching delay for
connections switching from FACH to DCH and vice-versa is not very
significant as compared to the transmission time of packets on the
FACH channel, given the fact that FACH is a low bandwidth channel
with high priority signaling traffic on it. We also argue that it
is advantageous for short flows to have small response times. This
observation and argument motivates us to propose the design of a new
cross-layer policy that we call FS-DCH ({\it at-least} flow-size
threshold on DCH) policy in which we try to achieve better response
time and slowdown for short flows. In
Section~\ref{network-model} we describe the network model and
simulation set-up that we have used for performance evaluation
of all the policies. Section~\ref{evaluation} leads to discussion on
the various observations that can be made from simulation graphs
obtained. We finally conclude in Section~\ref{conclusion}.

\subsection{Related Work}
\label{related-work}

Most of the existing channel switching policies are very simple,
timer and threshold based policies and do not involve any complex or cross-layer
switching criteria. Queue size threshold based policies have been
proposed in \cite{prabhu} in which a new connection is initially
allocated to FACH. On indication that the current flow of the
connection might be long (i.e., a long buffer queue for that source
is observed), then beyond some upper threshold, the Packet Scheduler
in UMTS tries to allocate a DCH to that connection (if one is
available). While on DCH, when the queue size of the connection
falls below another lower threshold, the connection is switched back
to FACH. The authors in \cite{prabhu} also present a modified threshold policy, in
which, while a connection is on DCH, if its queue size falls below a
lower threshold, a timer is started and the connection remains on
DCH. If there are no arrivals during the timer period, the
connection is switched back to FACH. The timer is used to let the
TCP acknowledgements (ACKs) reach the sender and release new packets. In
\cite{anderson}, the switching policy switches connections from FACH
to DCH when the number of packets transmitted (i.e., flow size) for a
given user on FACH exceeds a threshold. The choice of the threshold
depends on the load on FACH and other QoS conditions. In
\cite{qualcomm}, a switching policy based on bandwidth demand has
been proposed. A connection is switched from FACH to DCH if its
bandwidth demand exceeds a threshold and remains on FACH otherwise.
The channel switching schemes in \cite{ohta} work with blocking and
unblocking packets present in the RLC (Radio Link Control) and MAC sub-layers and
different schemes propose to transmit the unblocked packets on
either common or dedicated channels, differently.

The authors in \cite{prabhu} have implemented a PS+FCFS
queue system, i.e., the FACH channel uses a processor sharing (PS) scheduling mechanism
and the DCH channel uses an FCFS scheduling mechanism.
However, the policies that we propose in this paper are different from those
proposed by the authors in \cite{prabhu}, \cite{qualcomm}
and \cite{floberg}, since we use PS+Priority and LAS\footnote{LAS: Scheduling based on {\em Least Attained Service}}+Priority
queue systems for FACH and DCH channels and not PS+FCFS queue systems.
This will become more clear in the following section.

\section{Basic Channel Switching Policies}
\label{policy}

We propose three new basic threshold-based channel switching policies
for packet data transmission on the downlink of a single UMTS cell. They are
based on an extension of the modified threshold policy in \cite{prabhu}.
In all these policies, the FACH channel is served with either a PS or a LAS scheduling
mechanism and the DCH channel is implemented as {\em Priority scheduling}
with priority given to connections having maximum queue lengths.
Before we discuss in detail about the three channel switching policies,
we define below the notations used in their formal definitions:

\begin{itemize}
\item Let $Q(i)$ denote the queue length of a connection $i$ at the UMTS base station (NodeB).
\item Let $T_h$ and $T_l$ $(T_h \ge T_l)$ denote two thresholds on the queue length $Q(i)$
when the connection is on FACH and DCH channels, respectively.
\item Let $f(i)$ denote the cumulative flow size (i.e., number of packets
transmitted) over the FACH and DCH channels, for the current flow of a connection $i$.
\item Let `$s$' denote a threshold on the cumulative flow size $f(i)$ of the
current flow.
\item In all the policies described in this paper, a connection starts on FACH by
default and then if a DCH is available, it is switched to DCH
depending on different thresholds. If a DCH is not available then a request $r_i$
corresponding to this connection $i$ is added to a request set so that later
when a DCH is available, connection $i$ will be switched to DCH. Let $R$ denote
this request set.
\item Let $W(i)$ denote the total time for which a request $r_i$ of connection $i$
remains unserved. Alternatively, it denotes the total time for which a connection $i$
has been waiting to be switched to DCH since its request $r_i$ to switch to DCH
was added to $R$.
\item Let $N_{dch}$ denote the total number of DCH channels in a single UMTS cell.
\item Let $U_{dch}$ denote the total number of DCH channels that have been
allocated or currently in use in the UMTS cell. Note that $U_{dch} \le N_{dch}$.
\end{itemize}

\subsection{QS Policy}
\label{qs-policy}
In the QS (Queue Size) policy with parameter $T_h$, a new connection $i$ starts on the FACH channel and
waits for its queue length to
exceed an upper threshold $T_h$ before switching to DCH.
If there is no DCH channel available then a request $r_i$ for this connection
to switch to DCH is made. For a connection $j$ on DCH
when its queue length drops below the lower threshold $T_l$, a timer is started
for $T_{out}$ seconds. If there are packet arrivals
during the timer period, the timer is reset. When the timer expires, if 
the queue length of connection $j$ is still below
the lower threshold and another set of connections on FACH
are attempting to switch to DCH and no more free DCH channels are available, the
connection $j$ switches back to FACH.
Once this connection switches to FACH after a switch delay (of around 250ms),
a connection having the {\it maximum} queue length among the set of connections
on FACH that were attempting to switch to DCH, is switched to DCH.
In this way we give priority to the connections with the maximum
queue lengths while switching from FACH to DCH. This is what we mean
by Priority scheduling on the DCH channel. PS+Priority then implies that, FACH
uses PS scheduling mechanism and DCH uses Priority scheduling.
This PS+Priority queue system is the essential difference between our new basic
QS, FS and QSFS policies and the policies proposed
in \cite{prabhu} which use PS+FCFS queueing. We will see later
in Section \ref{evaluation} that
our new policies significantly improve over the modified threshold policy
in \cite{prabhu} by around $17\%$ in response time metrics. This leads
to the conclusion that PS+Priority queueing system is the main feature
due to which our new policies improve over the modified threshold policy
in \cite{prabhu}. The QS policy can be formally defined as follows:
\mybox{12cm}{
{\bf QS policy:}
The QS (Queue Size) policy is characterized by the following set
of rules:
\begin{itemize}
  \item A connection $i$ starts on FACH by default. It switches
to DCH if $Q(i) > T_h$  $\mathcal{\&}$  $U_{dch} < N_{dch}$.

If $Q(i) > T_h$  $\mathcal{\&}$  $U_{dch} = N_{dch}$  then $r_i$ is added to $R$.

  \item If connection $i$ is on DCH then if $Q(i) < T_l$, a timer is
started for duration $T_{out}$ seconds. If there are packet arrivals
during the timer period, the timer is reset. When the timer expires
and $Q(i) < T_l$, then, if (1) $U_{dch} = N_{dch}$  $\mathcal{\&}$  $R
\ne \phi$ then connection $i$ switches to FACH and connection $j$ with $r_j\in R$
switches to DCH, where connection $j$ is chosen such that
\[
r_j = \arg\max_{\substack{r_k \in R}} {Q(k)},
\]
else, (2) the connection $i$ remains on DCH and another timer of duration $T_{out}$
seconds is started.
\end{itemize}
}

In the above definition, once connection $j$ switches to DCH successfully, $r_j$ is
deleted from the request set $R$.
The motivation behind QS policy is to treat short flows and long
flows differently. The size of a flow can be
estimated by its queue size. Short flows will not exceed a
sufficient upper threshold $T_h$ on the queue size and will get
served on FACH. Thus, the idea is to avoid switching cost for short
flows as the cost may be more or comparable to the service
requirement of the short flows. Large-sized or long flows on the other hand
will see their buffer queue build-up and will be switched to DCH. 
An important advantage of this policy is that using only local information (i.e., queue size)
which is easily available, implicit queue size based scheduling can
be implemented in a scalable (with number of users) fashion.

\subsection{FS Policy}
\label{fs-policy}
In the FS (Flow Size) policy with parameter `$s$', the Packet Scheduler waits for the number of
packets served for the current flow of a connection to exceed a threshold `$s$' before
switching it to DCH. A connection on DCH switches back to FACH according
to the same rule as in QS policy. The FS policy can be formally defined as follows:

\mybox{12cm}{
{\bf FS policy:}
The FS (Flow Size) policy is similar to the QS policy except
for the fact that a flow size threshold `$s$' is used instead of the
queue size threshold $T_h$ on FACH. It is thus characterized by the
following set of rules:
\begin{itemize}
  \item A connection $i$ starts on FACH by default. It switches
to DCH if $f(i) > s$  $\mathcal{\&}$  $U_{dch} < N_{dch}$.

If $f(i) > s$  $\mathcal{\&}$  $U_{dch} = N_{dch}$  then $r_i$ is added to $R$.

  \item If connection $i$ is on DCH then it follows the same rule as in QS policy.
When connection $i$ switches to FACH successfully, $f(i)$ is set to $0$.
\end{itemize}
}

The FS policy is similar to QS policy except for the fact that the flow size is
directly computed from the number of packets served. A flow gets threshold
amount of service on FACH, exceeding which the flow is termed as a long flow
and switched to DCH. The policy is scalable with number of users
as the size of a flow can be computed locally.

\subsection{QSFS Policy}
\label{qsfs-policy}

In QSFS (QS \& FS combined) policy a connection
on FACH switches to DCH when conditions of both QS and FS policy are satisfied.
A connection on DCH switches back to FACH according to the same rule as in QS policy.
The QSFS policy can be formally defined as follows:

\mybox{12cm}{
{\bf QSFS policy:} In this policy, we combine
the QS and FS policies. It is thus characterized by the following
rules:
\begin{itemize}
  \item A connection $i$ starts on FACH by default. It switches
to DCH if $Q(i) > T_h$  $\mathcal{\&}$  $f(i) > s$  $\mathcal{\&}$  $U_{dch} < N_{dch}$.

If $Q(i) > T_h$  $\mathcal{\&}$  $f(i) > s$  $\mathcal{\&}$  $U_{dch} = N_{dch}$  then
$r_i$ is added to $R$.

  \item If connection $i$ is on DCH then it follows the same rule as in QS policy.
When connection $i$ switches to FACH successfully, $f(i)$ is set to $0$.
\end{itemize}
}

We defer the performance evaluation through simulations of the
above mentioned policies to Section~\ref{evaluation}.

\section{Designing a New Cross-Layer Channel Switching Policy}
\label{newpolicy}

Most of the software applications running over UTRAN use TCP as the transmission
protocol. TCP reacts to congestion and losses either by drastically reducing its congestion window size
after a timeout, or with some fluidity through fast retransmit procedures.
For short flows with small number of packets, a loss of one of the
last few packets is often detected only after a timeout, due to insufficient
NACKs received by the sender. Thus timeouts of short flows are not very
effective in reducing network congestion and
one of the most important aspects on the downlink channel is to sustain
efficient TCP performance by preventing timeouts of short flows and congestion in buffer
queues \cite{urtzi}. For example in peer-to-peer file exchanges, two
users exchange a small number of packets (generating
short flows) before one of them downloads a long heavy data file. Same is
true for FTP and HTTP web browsing traffic where packet exchanges between applications
running across UTRAN and a UE consist either entirely of short flows (if 
caching is enabled in the browser) or of short flows followed by a
long file transfer (if caching is not enabled). Similarly,
short flows are also generated by conversational voice packet
transfers (not streaming voice) where maximum
acceptable end-to-end delay according to the human perception
is around 400 ms. Thus from user ergonomics
point of view, it would seem advantageous to minimize the transfer times of
short flows by giving them priority over long flows and serving them
on a faster link \cite{urtzi}. This motivates us to design a cross-layer channel switching
policy in which the initial packets of a TCP flow are given priority
on a fast link and if this flow turns out to be a long flow then it can be
afforded to serve this flow on a slow link unless it builds up a very
large queue length on the slow link, in which case this long flow needs
to be switched back to the fast link.

In all the existing and basic channel switching policies discussed
previously in Sections \ref{related-work} and \ref{policy}, respectively,
a new flow of a connection always starts on the slow FACH channel and waits until
some threshold parameter has been attained, before switching to the fast DCH channel.
For short data bursts of say less than 10 packets,
it may take a long time (on slow FACH) for them to surpass any threshold parameter
or they may never surpass it at all (due to insufficient number of packets). On
the other hand, long flows with a large number of packets will most probably
surpass the thresholds and get a chance to be transmitted on the fast DCH channel.
Thus there is a possibility that short flows in their entirety will suffer high
transmission times on the slow FACH channel, where as for long flows even though
their initial few packets are transmitted on the FACH channel, their overall
transmission time may improve since most of their (remaining) packets are
transmitted on the DCH channel. This intuition can be further strengthened
by some concrete calculations that follow.

Let us take a closer look on the FACH channel. The FACH channel has a very low
set-up time, usually has a capacity of around $33$ kbps and has a high priority
signal traffic (from a constant bit rate (CBR) source) running on it apart
from the data packets. The CBR source transmits signal traffic at the rate
of around $24$ kbps. So a short data burst of say $10$ packets of $1$ kbyte each will
take approximately $8.88$ seconds (or $2.42$ seconds in the best case when CBR
traffic is absent) to be transmitted on the FACH channel. Now let us consider
the DCH channel. The DCH channel has a capacity of around $384$ kbps. There is a
set-up time of around $250$ ms for the DCH channel which is much higher than the
set-up time of the FACH channel. So unlike the mechanism used in existing and basic
switching policies, if a connection starts on FACH and switches to DCH
immediately without waiting to attain any threshold, a $10$ kbytes burst will get
transmitted in approximately $0.25 + 10\times 8/384 = 0.25 + 0.208 = 0.458$ seconds.
This significantly reduces the transmission time by a factor of about $20$ in
the presence of CBR traffic and about $5$ in its absence. Thus, switching a
new flow to DCH as soon as it starts can be beneficial for short data bursts
which would have otherwise suffered high transmission times on the slow FACH
channel. This clearly illustrates that the existing and basic policies discussed previously
in Sections \ref{related-work} and \ref{policy}, respectively,
suffer from a major drawback. The drawback being that
a new flow is allowed to transmit initially on slow FACH for a long time
(by the threshold mechanism) before it gets a chance to be transmitted on
the fast DCH.

The above argument gives us the motivation to design a cross-layer channel
switching policy in which the initial few packets of a new TCP flow of a
connection on FACH are given priority on the fast DCH channel by
switching the connection from FACH to DCH {\it as soon as possible}.
If this new flow is a short flow then it will be entirely served on
DCH thus ensuring minimum transfer times for short flows, as
explained with the help of some calculations in the previous
paragraph. Otherwise if this flow turns out to be a long flow, then
later if the buffer queue length of the associated connection falls
below a threshold $T_l$, the connection is either {\it preempted}
and switched back to FACH to allow other new flows on FACH to switch to
DCH, or the connection remains on DCH and then ultimately times out
(in the absence of packet arrivals during an inactivity timer
period) and is switched to FACH indicating the end of current flow
on the connection. Thereafter, any new packet arrivals on this timed
out connection on FACH will be termed as a new flow. Thus at any given
instant there are either {\em new} flows on FACH attempting to switch to
DCH, or there are {\em old} flows on FACH (which may also be long with a
high probability) which have already transmitted their initial few
packets (say at least first `$s$' packets) on DCH. If the buffer queue
length of the connections with old flows surpasses the threshold
$T_h$, then they attempt to switch to DCH again in order to minimize
the use of FACH channel, since it is a very slow channel that can
cause significant increase in transmission times.

Note that in our new policy described above, a new connection must
always necessarily start transmitting on the common
FACH channel, since the number of DCH channels are interference limited and a DCH may not always be
available to be allocated for a new connection. When a connection $i$ on FACH attempts
to switch to DCH and if no DCH channel is available, a request $r_i$
to switch to DCH is pushed into a request set $R$  and this request is served when
a DCH channel is available later.

We call the strategy of allowing a new flow to transmit at least its first `$s$' packets
on DCH as the {\it first `$s$' on DCH} mechanism and it is one of the two key features of our new improved
switching policy. The other key feature is the use of {\it dual-level priority
switching} mechanism. This mechanism works as follows. If more than one
connections on FACH are candidates (i.e., they have requested to switch to DCH)
to be switched to a single available DCH channel, then the dual-level priority
switching mechanism chooses only one connection among all connections with new flows,
on a {\it first-come first-served} (FCFS) basis, to be switched to DCH. In the
absence of connections with new flows, the connection with the {\it maximum queue
length} among all connections with old flows, is switched to DCH. We term
our cross-layer channel switching policy as FS-DCH ({\it at-least} flow-size
threshold on DCH) policy and it can be formally defined as follows:

\mybox{12cm}{
{\bf FS-DCH policy:}
The FS-DCH policy is defined as follows:
\begin{itemize}
  \item A connection $i$ starts on FACH by default. A connection switches
to DCH if (1) $f(i) \le s$  $\mathcal{\&}$  $U_{dch} < N_{dch}$  or
(2) $f(i) > s$  $\mathcal{\&}$  $Q(i) > T_h$  $\mathcal{\&}$  $U_{dch} < N_{dch}$.

If (1) $f(i) \le s$  $\mathcal{\&}$  $U_{dch} = N_{dch}$ or
(2) $f(i) > s$  $\mathcal{\&}$  $Q(i) > T_h$  $\mathcal{\&}$  $U_{dch} = N_{dch}$
then $r_i$ is added to $R$.

In this rule, the condition (1) causes a new connection starting on FACH
to attempt to switch to DCH as soon as possible.

  \item If connection $i$ is on DCH and $Q(i) < T_l$ then
\begin{description}
\item[(a)]if $f(i) \le s$, then it follows the same rule as in QS policy.
When connection $i$ switches to FACH successfully, $f(i)$ is set to $0$.
\item[(b)]if $f(i) > s$, then

if (1) $U_{dch} = N_{dch}$  $\mathcal{\&}$  $R \ne \phi$,
connection $i$ is {\it preempted} and it switches to FACH and
connection $j$ with $r_j\in R$ switches to DCH, where 
connection $j$ is chosen such that $f(j) \le s$ (its a new flow) and
\[
r_j = \arg\max_{\substack{r_k \in R}} {W(k)}.
\]
If there is no such connection that satisfies the condition $f(j) \le s$
then connection $j$ is chosen such that $f(j) > s$ (its an old flow) and
\[
r_j = \arg\max_{\substack{r_k \in R}} {Q(k)},
\]
else, (2) it follows the same rule as in QS policy.

\end{description}
\end{itemize}
}

In the above definition, once connection $j$ switches to DCH successfully, $r_j$ is
deleted from the request set $R$. We defer the performance evaluation through simulations of the
FS-DCH policy to Section~\ref{evaluation}.

\section{UMTS Network Model \& Simulation Setup}
\label{network-model}

In this section we describe the UMTS network model
that we use for performance evaluation of the various aforementioned policies
through simulations. The model described here is very similar
to the one in \cite{prabhu}. We consider a network model with $N_{tcp}$ TCP sources
which need to send data to
mobile receivers. We assume a single cell scenario with one NodeB base
station and several mobile stations which act as destinations for
TCP traffic. The TCP sources are assumed to be connected to the
base station of the cell with a high speed (5mbps, 30ms) link.
The base station can transfer data from a TCP source on either
DCH or FACH at a given time. There is one FACH and $N_{dch}$ DCH channels
in the system. The FACH is a time division multiplexed channel.
In addition to any TCP connections which may be present on a FACH,
there is signaling traffic which must be transmitted on the FACH.
The signaling traffic has priority over the TCP connections. During
the silence periods of the signaling traffic, data from one or more
TCP connections can be transmitted on the FACH. Data
from the TCP connections is assumed to be transmitted on the FACH
with a PS or LAS service mechanism. If all the DCHs have
a TCP connection allocated, a connection on DCH should be first
switched to FACH before a connection from FACH can be switched on to
a particular DCH. This means that a switch can take up to $500$ms
(if there is already a TCP connection configured on the DCH and if
we consider the connection release time to be the same as the connection
set-up time). Switching from one channel to another is costly in time and
signaling. In the model we assume that there exists a queue
corresponding to each TCP connection in the NodeB base station. The
base station is hence able to track both the queue length and
the number of packets served (flow size) for each
connection. During the switching time from one
channel to another, no packets from the queue of the TCP
connection being switched can be transmitted. While a connection
is switching from one channel to another, the ACKs of a TCP connection
traverse the original channel until the switch is completed.



\begin{figure}
\centering\includegraphics[width=3.4in]{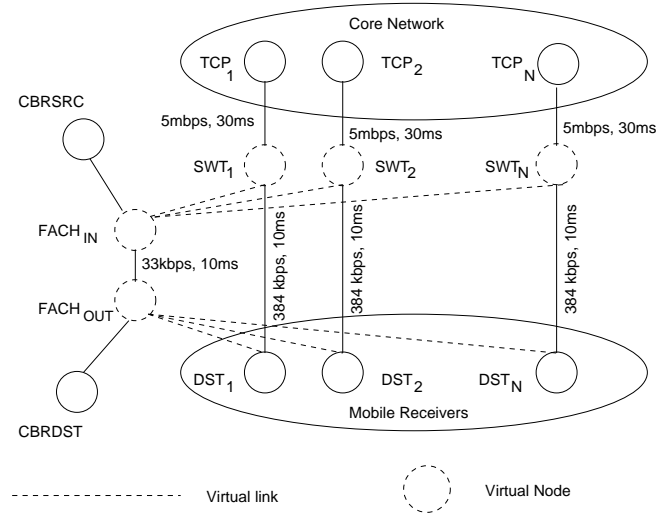}
\caption{Simulation Setup}
\label{setup}
\end{figure}
The simulation setup for the above described network model is presented in
Figure~\ref{setup}. Each TCP source node $TCP_i$ is connected to a
routing node called Switch ($SWT_i$). $SWT_i$ is present inside the
NodeB base station and can be connected either to the $FACH_{IN}$ or
directly to the TCP destination via the DCH. The $SWT_i$ node
has been introduced to simplify the simulations and may not be present inside a
real NodeB base station. The $FACH_{IN}$ is another virtual node which
simulates either the PS or LAS service discipline
taking place on the FACH. In the PS discipline, the
node $FACH_{IN}$ gives priority to the traffic from $CBRSRC$ while
serving the packets from the $SWT_i$'s (only those which are
currently not transmitting on DCH) in a round-robin manner. We note
that there are no queues at $FACH_{IN}$ and all the packets are
either queued at $SWT_i$ or at the $CBRSRC$. The $CBRSRC$ simulates
a constant bit rate source of signaling/control traffic. It
generates packets at rate $R_{sig}$ and is assumed to be present
within the NodeB. The signaling traffic flows from the NodeB
to the mobile receivers. Even though we model the
destination of the signaling traffic ($CBRDST$) as another node
different from the mobile destinations $DST_i$, we note that it does not affect the
simulations as simultaneous transfer of data and control packets to
the same mobile receiver is indeed possible in UMTS when different channels are
used. The links $SWT_i-FACH_{IN}$ are virtual
links within the base station and thus have zero delay. Note that
the data from $SWT_i$ to $DST_i$ can take two different routes i.e.,
$SWT_i-FACH_{IN}-FACH_{OUT}-DST_i$ (via FACH) or simply $SWT_i-DST_i$ (via
DCH). At any given time only one of the above two routes can be
active for a given connection. Although in the simulation scenario
we have as many DCH links as TCP source nodes, the simulation allows
us to activate not more than $N_{dch}$ DCH channels at a time, which
may be chosen strictly smaller than the number of TCP sources
($N_{tcp}$). In the simulations we switch over from FACH to DCH by
changing the cost of the links and recomputing the routes. This is
done as follows. Initially, the cost of direct path from the
Switch to the TCP destination is set to 10 and the cost of all other
links to 1.  Hence, the traffic gets routed through the FACH. When a
switch is required, the cost of DCH is set to 1 and the routes
are recomputed. This activates the DCH and the traffic gets routed
on the DCH. 

\subsection{Limitations and Assumptions}
\label{limitations}

The layer 2 in UTRAN consists of two sub-layers: MAC layer and RLC
(Radio Link Control) layer. As described previously, the 
physical layer (layer 1) offers services to the MAC layer via 
transport channels of two types: dedicated channels and common
channels. The MAC layer in turn offers services to the RLC layer
above it through logical channels. The different logical channels
are mapped to the transport channels in the MAC layer. The two most
important {\it logical entities} in MAC layer are MAC-c/sh and
MAC-d. The MAC-c/sh entity handles data for the common and shared
channels, where as the MAC-d entity is responsible for handling data
for the dedicated channels. However, the execution of switching
between common and dedicated channels is also performed by the MAC-d
entity in UTRAN (in the serving RNC) based on a switching decision
derived by the channel switching algorithm that resides in the RRC
(Radio Resource Controller).

Data packets or SDUs (Service Data Units) arriving from upper layers
are segmented into smaller data packets or PDUs (Protocol
Data Units) by the RLC layer and PDUs are then forwarded to the MAC layer.
In our network model used to carry out the simulations for performance
evaluation of various switching policies, we do not consider the segmentation
of SDUs into PDUs. In other words, we do not model the RLC layer since the
main focus of this paper is to investigate the channel switching
mechanism. We thus model only the MAC-d entity in the MAC layer.
We also do not take care of packet loss, mobility and handovers, since considering them
would highly complicate the model and it is beyond the scope of this paper.

\subsection{Simulation Parameters}

We use ns-2 \cite{ns2} in order to simulate the various switching
policies for performance evaluation. The simulation parameters used
are described below:
\begin{itemize}
    \item We consider 2 cases for the number of dedicated channels, $N_{dch}$
    equal to 1 and 2. We vary the number of TCP sources as
    $N_{tcp}=2,3,5$ and $10$.
    \item The duration of simulations is taken to be $200,000$ secs.
    in order to reach stationarity. 
    \item The transmission rates for FACH and DCH channels are considered to be $33$ kbps
    and $384$ kbps, respectively.
    \item The switching cost $D_{sw}$ (in terms of time) between FACH and DCH channels
    is $250$ms.
    \item We consider the signaling traffic source (non TCP traffic source)
    that uses the FACH, to be a constant bit rate CBR source with
    rate $R_{sig} =24$ kbps. It sends a $1$ kbyte packet at an interval
    of $1/3$s and has a non preemptive priority over TCP traffic.
    \item The TCP connection traffic model is as follows: In a
    TCP connection, data arrives in bursts. The number of
    packets in a burst has a Pareto distribution and the shape
    parameter is taken to be $k=1.1$. The
    average file size is taken to be $FS_{avg}=30$ kbytes. 
    A TCP connection alternates between ``ON'' and ``OFF'' states.
    The ON state is comprised of several bursts and no packets are
    transmitted during the OFF state.
    In the ON state, the inter-arrival time between successive
    bursts is exponentially distributed with mean $T_{ON}=0.3$s.
    At the end of each burst in ON state, the connection goes into OFF
    state with probability $P_{OFF}=0.33$. It remains in the OFF state
    for an exponentially distributed
    duration with mean $T_{OFF}=5$s before it goes back into
    ON state again.
    \item The value of $T_l$ (lower threshold on DCH) is taken as $1$ and the
    packet size as $280$ bytes.

\end{itemize}

\section{Performance Evaluation of Policies}
\label{evaluation}

In this section, we analyze the results obtained from an extensive
set of simulations of the various channel switching policies that we have
discussed until now. We study PS scheduling of TCP sources on the
FACH channel for QS, FS, QSFS and FS-DCH policies. In addition to
this we also study LAS scheduling on FACH channel for FS policy
specifically. LAS scheduling can also be studied with other
policies, but since LAS looks at the number of served packets, which
relates to the flow size, FS policy is the most appropriate one to
study with LAS scheduling.

\begin{figure}
\begin{minipage}{2.8in}
\centering
\includegraphics[width=2.6in]{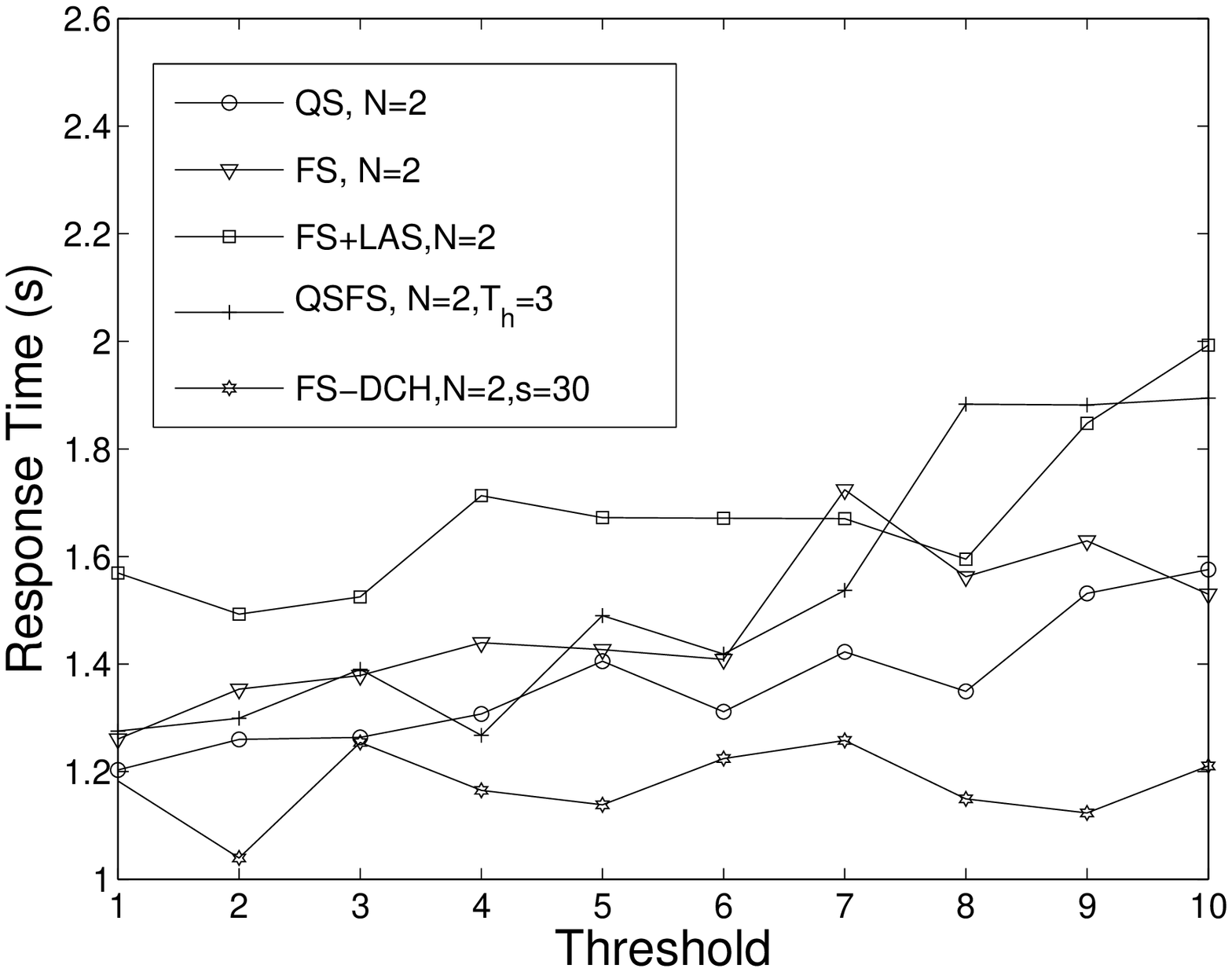}
\centerline{(a)}
\end{minipage}
\begin{minipage}{2.8in}
\centering\includegraphics[width=2.6in]{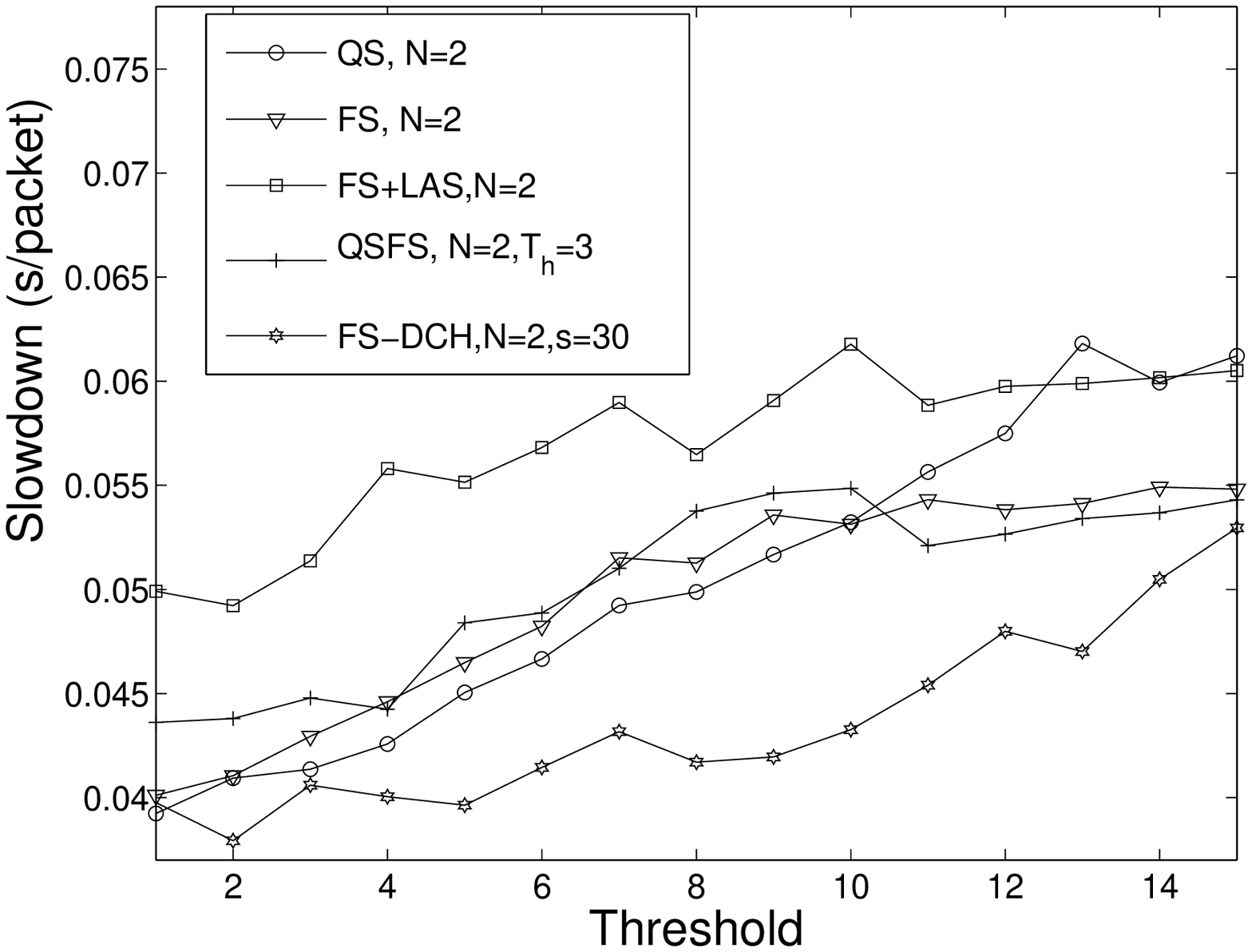}
\centerline{(b)}
\end{minipage}
\caption{Comparison of different policies in terms of response time
and slowdown metrics for $N_{tcp}=2$, $FS_{avg}=30$ kbytes and $N_{dch}=1$.}
\label{N2}
\end{figure}

\begin{figure}
\begin{minipage}{2.8in}
\centering\includegraphics[width=2.6in]{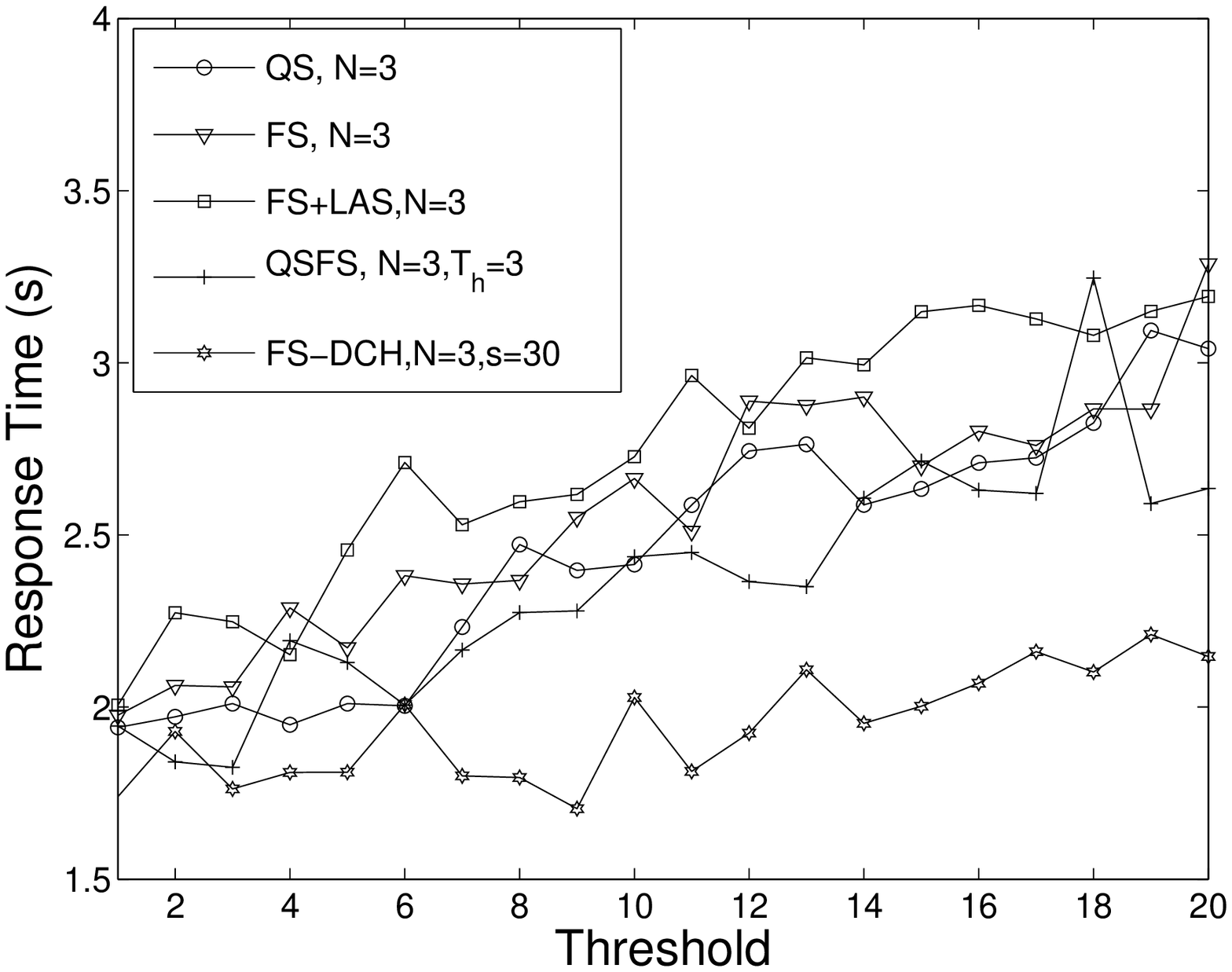}
\centerline{(a)}
\end{minipage}
\begin{minipage}{2.8in}
\centering\includegraphics[width=2.6in]{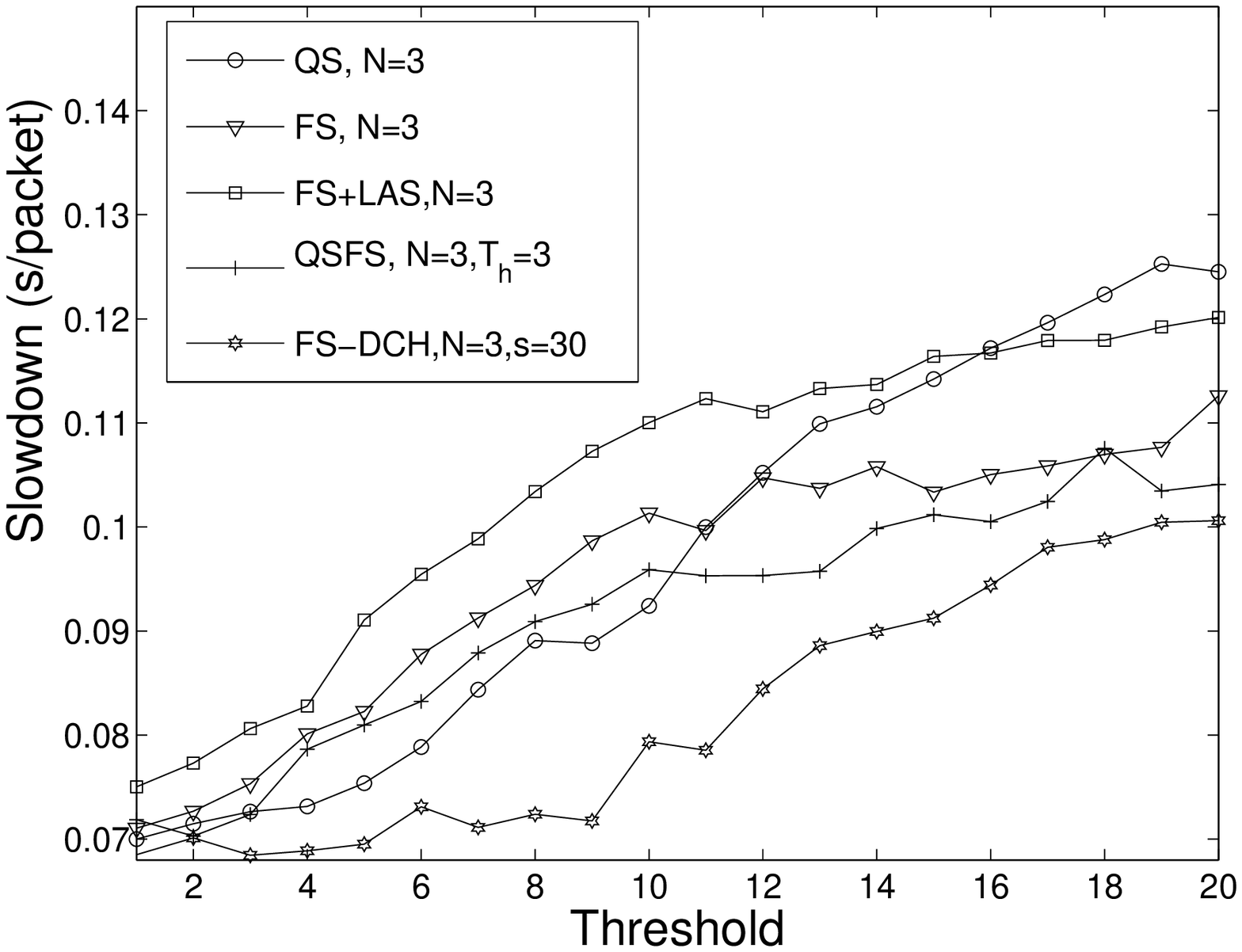}
\centerline{(b)}
\end{minipage}
\caption{Comparison of different policies in terms of response time
and slowdown metrics for $N_{tcp}=3$, $FS_{avg}=30$ kbytes and $N_{dch}=1$.}
\label{N3}
\end{figure}

\begin{figure}
\begin{minipage}{2.8in}
\centering\includegraphics[width=2.6in]{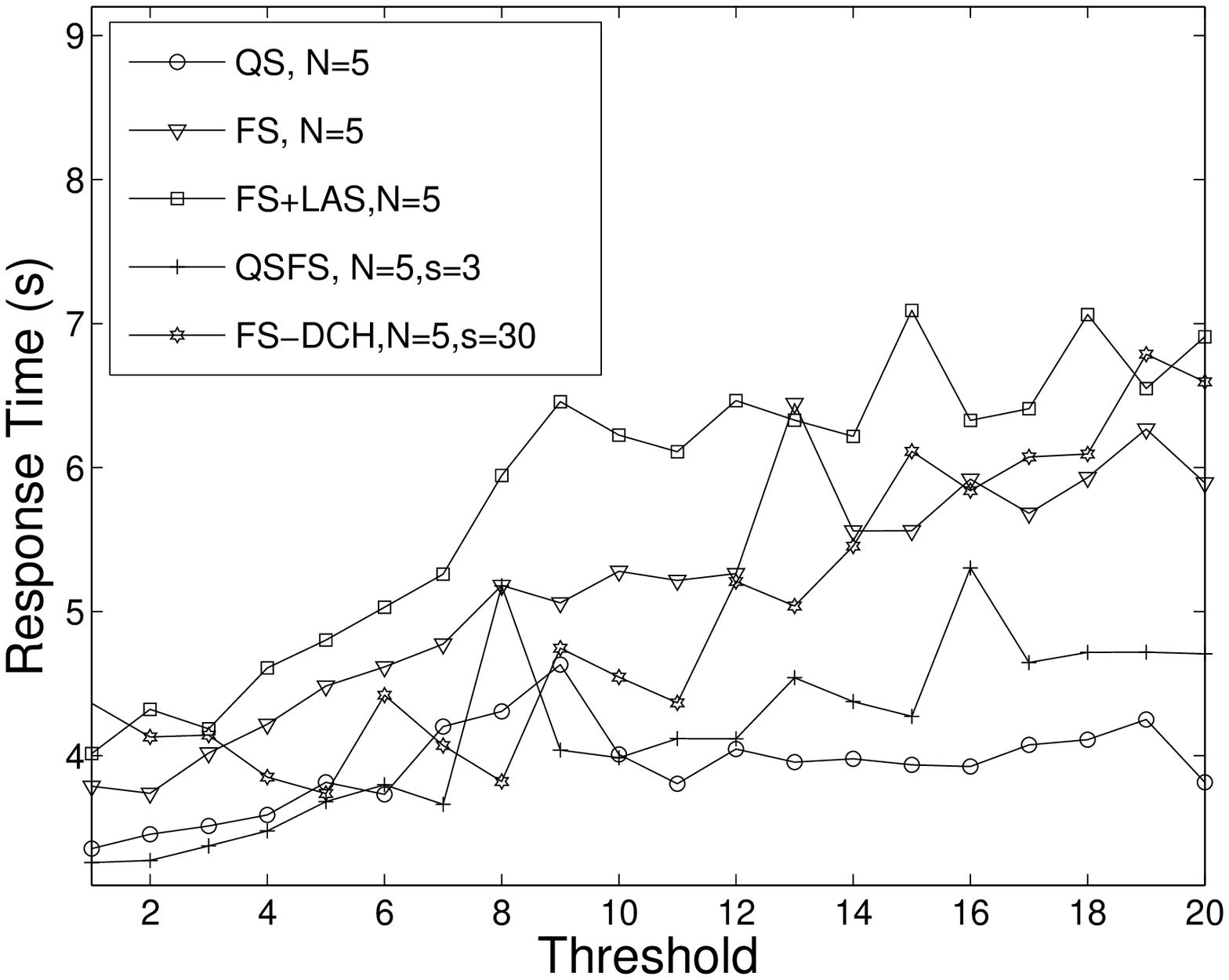}
\centerline{(a)}
\end{minipage}
\begin{minipage}{2.8in}
\centering\includegraphics[width=2.6in]{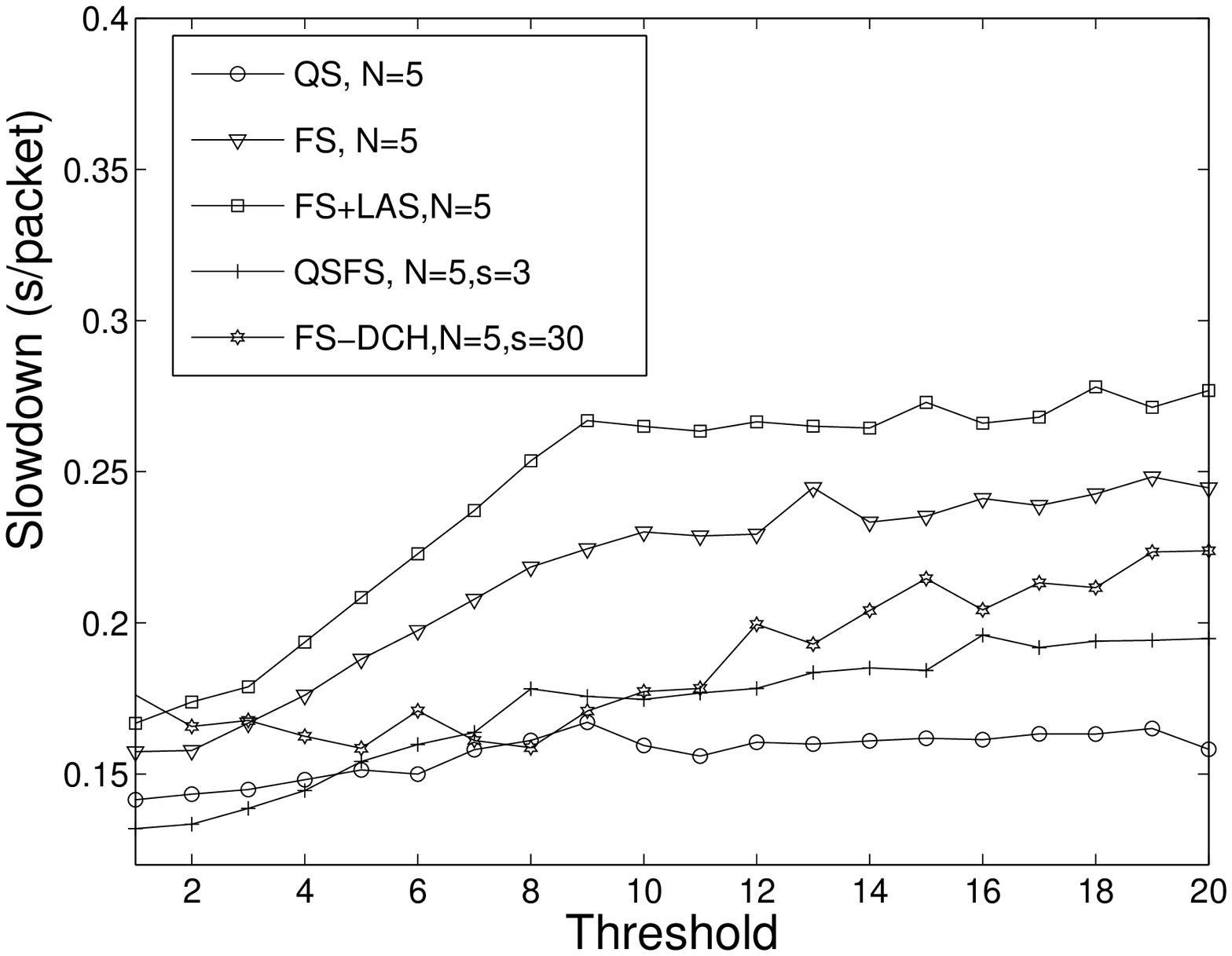}
\centerline{(b)}
\end{minipage}
\caption{Comparison of different policies in terms of response time
and slowdown metrics for $N_{tcp}=5$, $FS_{avg}=30$ kbytes and $N_{dch}=1$.}
\label{N5}
\end{figure}

\begin{figure}
\begin{minipage}{2.8in}
\centering\includegraphics[width=2.6in]{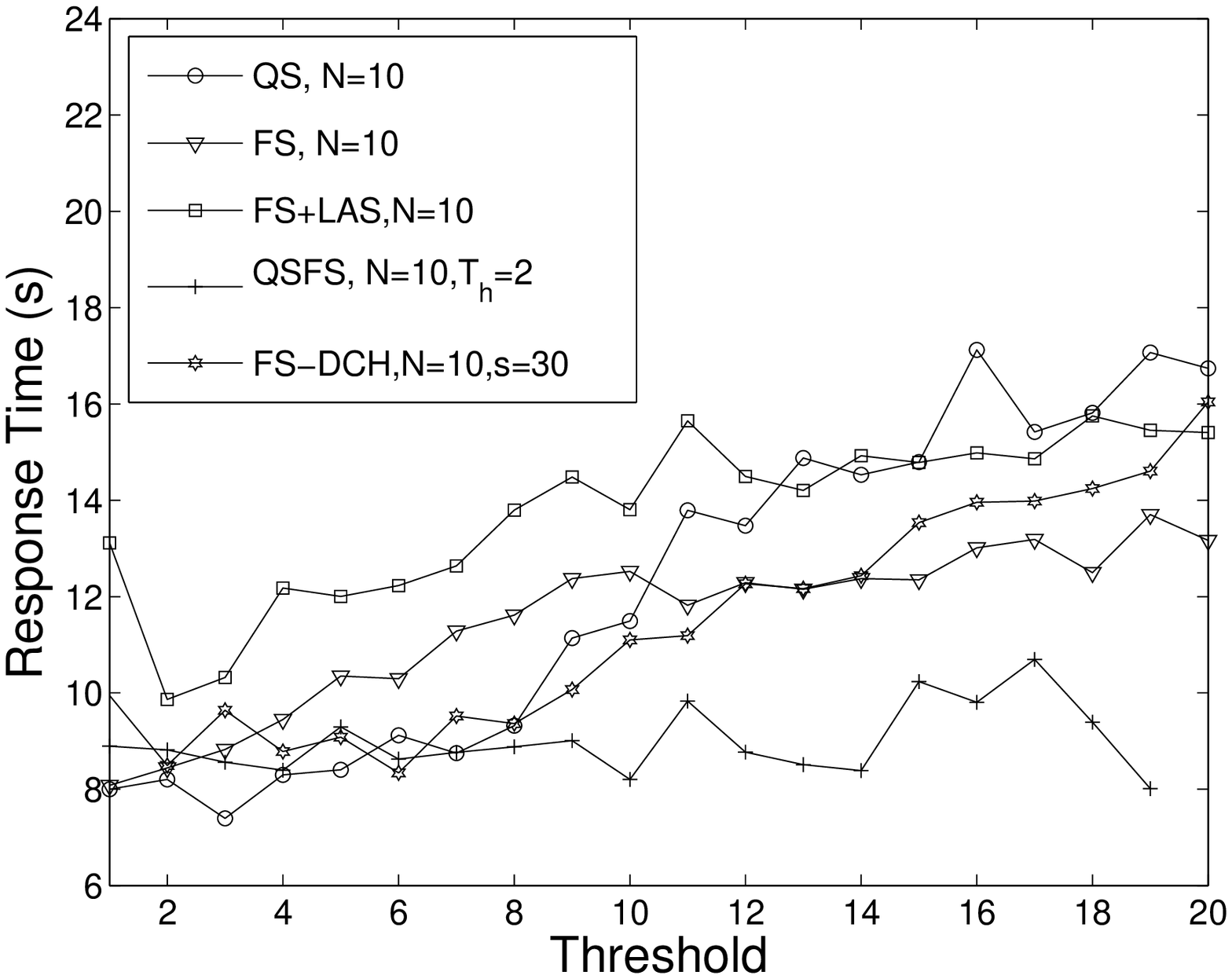}
\centerline{(a)}
\end{minipage}
\begin{minipage}{2.8in}
\centering\includegraphics[width=2.6in]{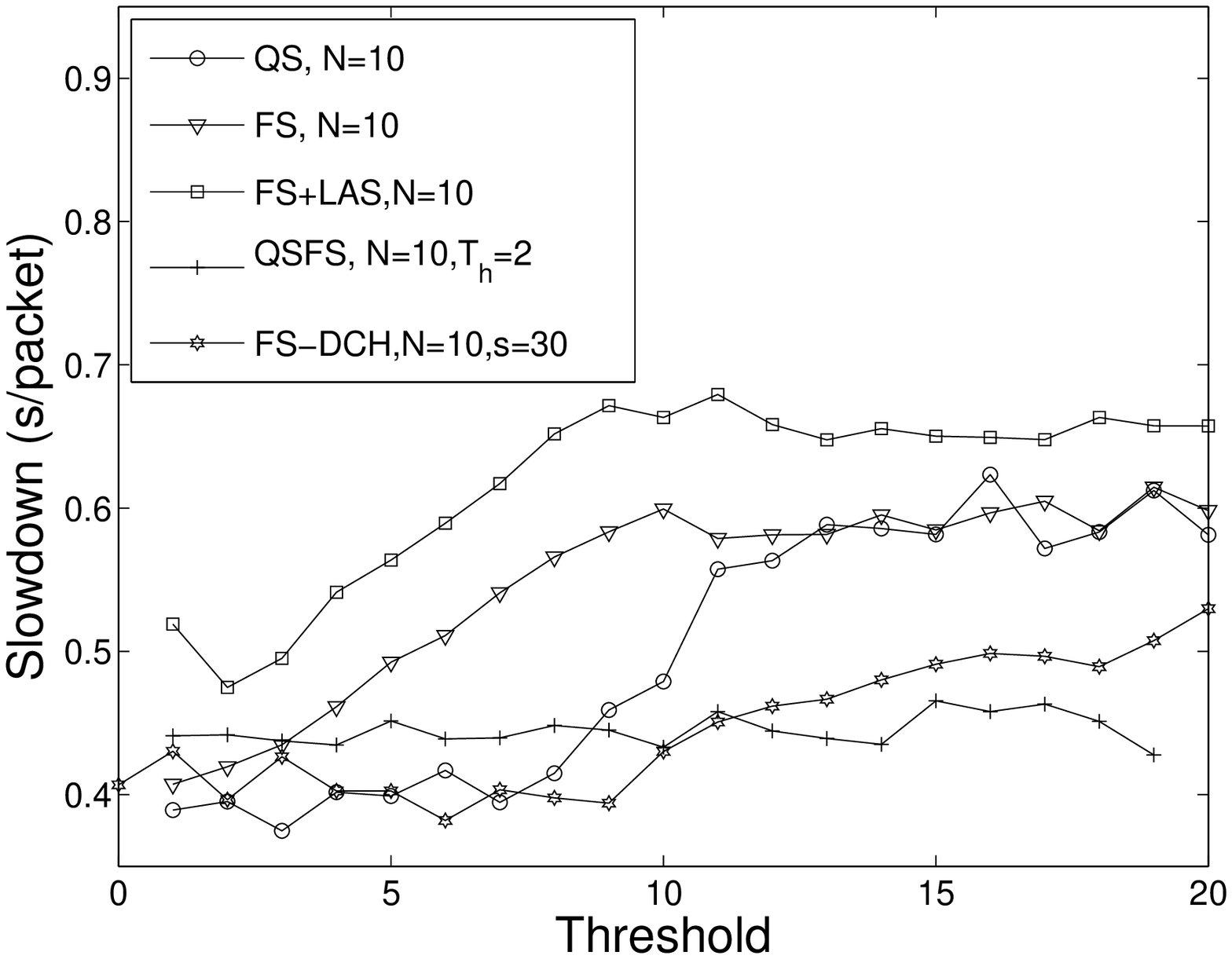}
\centerline{(b)}
\end{minipage}
\caption{Comparison of different policies in terms of response time
and slowdown metrics for $N_{tcp}=10$, $FS_{avg}=30$ kbytes and $N_{dch}=1$.}
\label{N10}
\end{figure}

In Figures \ref{N2}-\ref{N10}, we compare different policies in terms of
response time and slowdown as a function of $s$ or
$T_h$, as the case may be. The response time is calculated as the total
average time required to completely transmit a burst. By completely transmitting
a burst, we mean the time until a TCP ACK for the last packet of a burst sent, is received
at the sender side. Slowdown is defined as the response
time divided by the average burst size. So, for an average burst size of $x$, if $T(x)$ is
its response time then the slowdown $S(x)$ is defined as $\frac{T(x)}{x}$.
In Figure \ref{N2}(a), we observe that FS-DCH outperforms all other policies in
terms of response time, where as FS+LAS scheme has the highest
response time. The other three schemes have comparable response
times. The average improvement in response time achieved by FS-DCH over
all other policies is around $30\%$.
Within the range of threshold values shown,
we observe an increasing trend in response time under
all policies except for FS-DCH. The QS policy performs slightly better
than the FS policy in minimizing mean response time.
\junk{
On one hand, under FS policy larger threshold can ensure that short
flows can finish on FACH without having to switch to DCH
(but it also implies that many flows, including long flows, tend
to stay on FACH for a long increasing period). On the other hand,
small threshold helps long flows switch to DCHs early without
affecting short flows much.}

Under QS, FS and QSFS policies, at higher values of $T_h$ an increase in the response
time is observed because a higher value of $T_h$ implies more time
is spent in the FACH. The FACH is a low bandwidth channel which
has high priority signaling traffic on it. This results in low average
bandwidth being shared amongst the TCP connections due to the following reason. 
For a TCP connection, the switch to DCH is based on its current buffer size which in turn
depends on its current congestion window size. The congestion window size is incremented
whenever an ACK is received by the sender. When a TCP connection is on a low
bandwidth link, the window builds up slowly due to delay in receiving an ACK.
This slow buildup of the window size results in slow buildup of the current
buffer size. As the value of $T_h$ is increased, a TCP connection has to spend
more time on the slow FACH, resulting in a higher delay.

The comparison of average slowdown in Figure~\ref{N2}(b) shows that
the slowdown metric follows the same trend as that of average response time.
FS-LAS has the highest slowdown and FS-DCH has the lowest. Other policies
perform almost the same except that performance of QS worsens for higher
values of the threshold.

From the above discussions it can be concluded that the cross-layer FS-DCH policy is
better than all other policies for $Ntcp=2$.

In Figure~\ref{N3}, we plot the average response time and
slowdown for $N_{tcp}=3$. It can be easily seen that FS-DCH again performs the
best in terms of both response time and slowdown and all other policies
perform comparably among themselves. The average improvement in response time achieved by FS-DCH over
all other policies is around $36\%$.

In Figure~\ref{N5}, we plot the average response time
and slowdown for different policies for $N_{tcp}=5$, where we can
easily see that the performance metrics are very different from
those observed for $N_{tcp}=2$ and $3$.
Here, on an average, QS achieves the lowest response time and slowdown instead of FS-DCH.
The improvement of QS over all other policies in response time metrics is
on an average $18\%$.
FS-LAS gives the highest response time and slowdown and FS-DCH is
worse than QSFS in terms of both response time and slowdown.

In Figure~\ref{N10}, for $N_{tcp}=10$, we observe that on an average, QSFS policy performs
the best in terms of both response time and slowdown. The average improvement of QSFS
policy over other policies in response time metrics is around $21\%$.
For threshold values $>12$ packets, the response times of QS and
FS+LAS are comparable.

As mentioned before, our QS, FS and QSFS policies are based on the
modified threshold policy proposed in \cite{prabhu}, but are different
since we use PS+Priority queueing instead of PS+FCFS queueing.
If we compare the simulation results of our new basic QS, FS and QSFS policies
with results of the modified threshold policy proposed in \cite{prabhu}
we can easily observe that our new switching policies improve on the modified
threshold policy by around $17\%$ in terms of response time.

\subsection{Effect of Number of DCHs}

\begin{figure}
\begin{minipage}{2.8in}
\centering\includegraphics[width=2.6in]{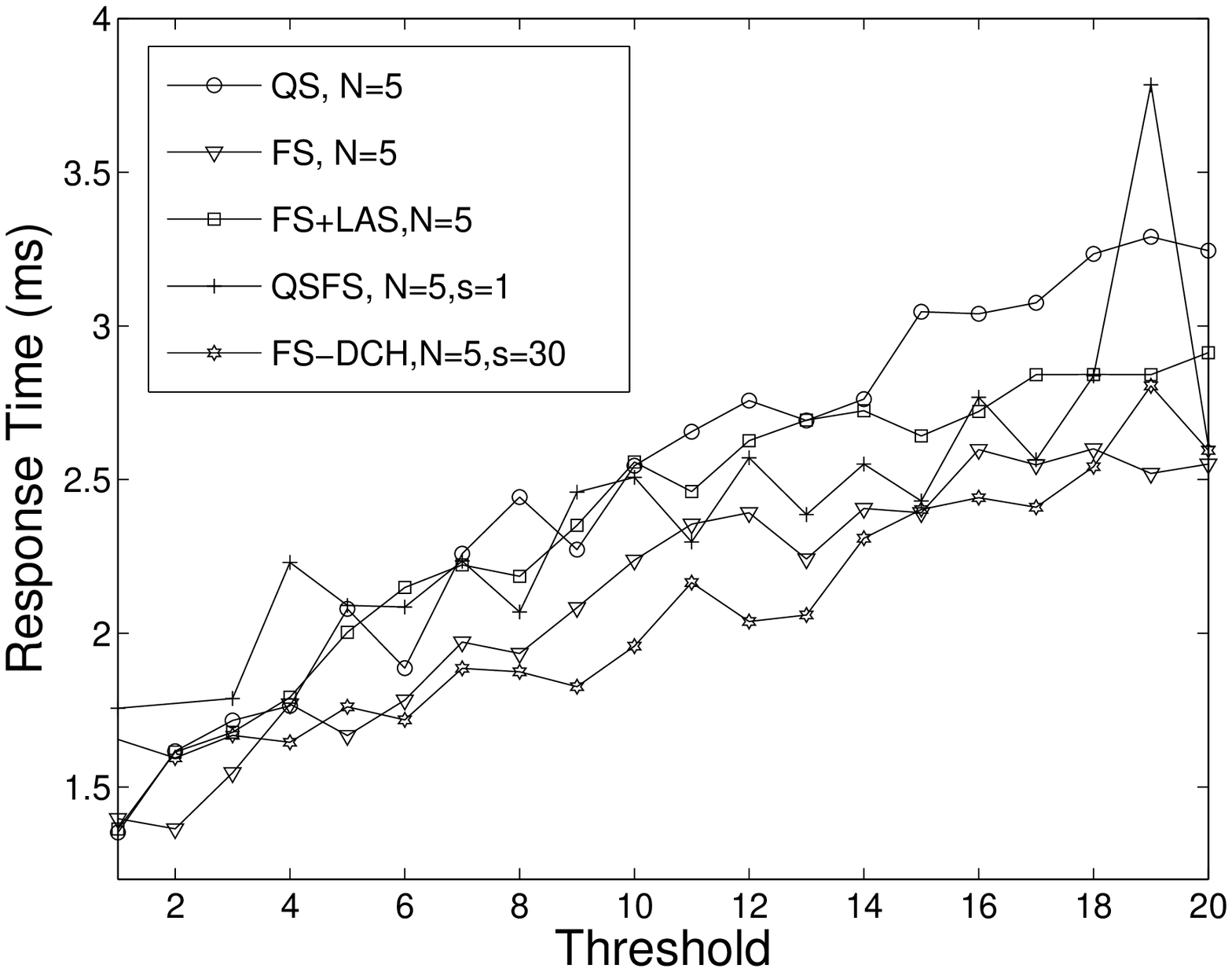}
\centerline{(a)}
\end{minipage}
\begin{minipage}{2.8in}
\centering\includegraphics[width=2.6in]{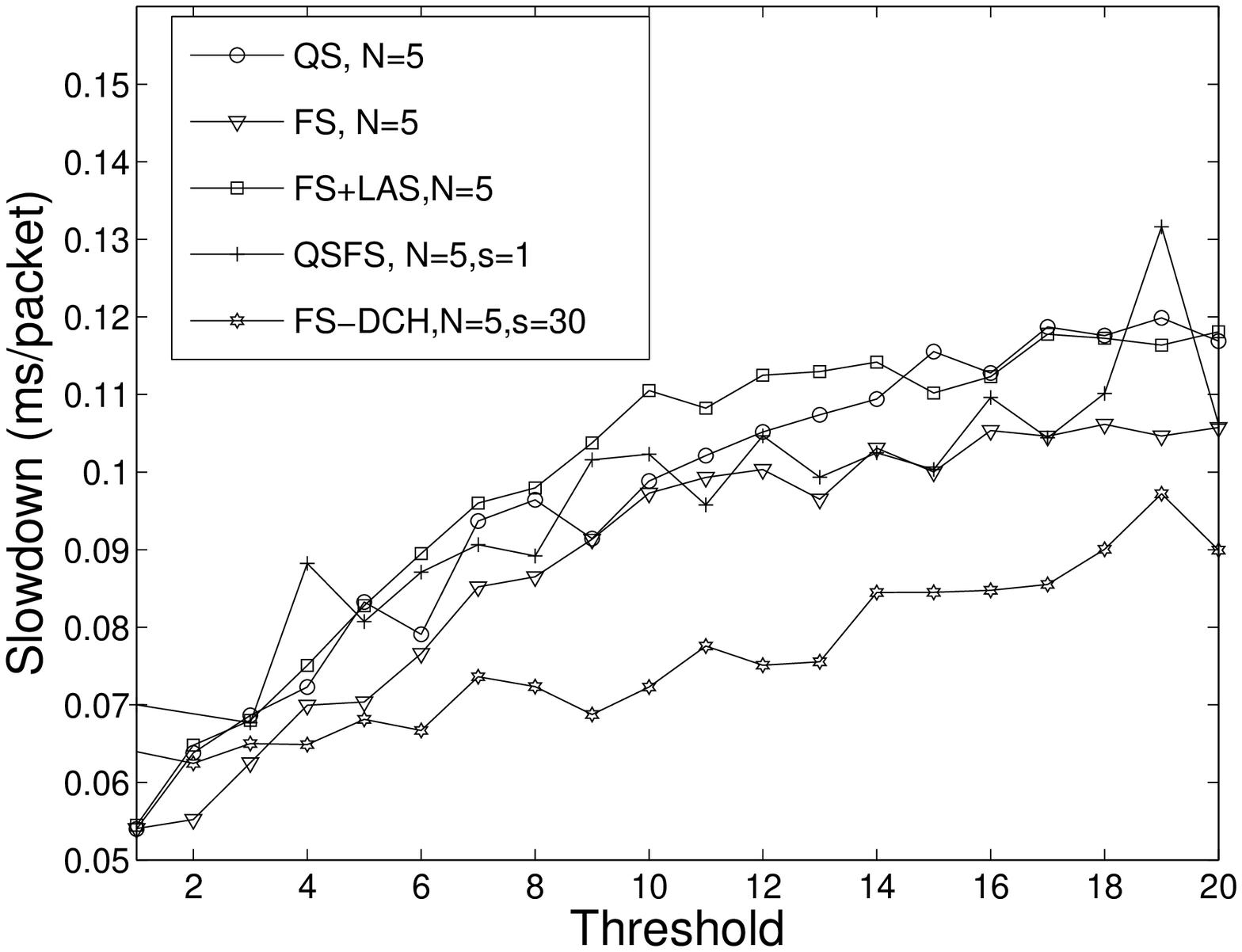}
\centerline{(b)}
\end{minipage}
\caption{Comparison of different policies in terms of response time
and slowdown metrics for $N_{tcp}=5$, $FS_{avg}=30$ kbytes and $N_{dch}=2$.}
\label{N5_server2}
\end{figure}

Here we discuss about performance of different policies
when the number of dedicated channels is increased to 2.
In Figure~\ref{N5_server2}(a) we observe that FS-DCH again outperforms
other policies in terms of response time. Other policies perform comparably.
In Figure~\ref{N5_server2}(b), we observe that slowdown follows the 
same trend as that of response time.

\section{Conclusion}
\label{conclusion}

In this paper, we have proposed several {\it scalable} channel 
switching policies for packet data transmission on UMTS downlink. 
The basic threshold based policies QS, FS and QSFS are based on an extension of the modified
threshold policy in \cite{prabhu}. In our policies, we use PS+Priority and
LAS+Priority queue systems for FACH and DCH channels instead of
PS+FCFS queue system used in \cite{prabhu}. Simulation results show
that our new basic switching policies improve on the modified
threshold policy in \cite{prabhu} by around $17\%$ in response time
metrics. We have further proposed a new and improved {\em cross-layer}
channel switching policy that we call FS-DCH ({\it at-least}
flow-size threshold on DCH) policy. FS-DCH is a biased policy 
that improves the performance of TCP flows by giving priority
to short flows on the fast DCH channel. Results obtained from
extensive simulations show that for a given simulation set-up 
of $N_{dch}=1$ and $FS_{avg}=30$ kbytes, FS-DCH performs better than the basic QS, FS
and QSFS policies, when the number of TCP connections $N_{tcp}$ is low.
For example for $N_{tcp}=2$ and $3$, FS-DCH gives a
significant average improvement of $30\%$ to $36\%$, respectively, over all other
policies, in terms of response time. However, with higher number of
TCP connections $N_{tcp}$, the best performing policies are the ones that
are queue size based i.e., QS and QSFS policies. They give an average
improvement in response time of $18\%$ and $21\%$, respectively. One can also
observe that the best performing policies in terms of response time,
also outperform in terms of slowdown. 

\begin{table}[htbp]
\begin{center}
\begin{tabular}{|l|c|c|} \hline
\em{$N_{tcp}$} &\em{Best Performing} & \em{Percentage gain} \\
& \em{Policy} & \em{in Response Time} \\ \hline
 2 & FS-DCH & $30\%$ \\
 3 & FS-DCH & $36\%$ \\
 5 & QS & $18\%$ \\
10 & QSFS & $21\%$ \\ \hline
\end{tabular}
\caption{Best performing policies for different values of $N_{tcp}$, $N_{dch}=1$ and $FS_{avg}=30$ kbytes}
\vspace{-0.2cm}
\end{center}
\label{arbit-label}
\end{table}

It can also be concluded that there is no single policy that can
be termed as an overall best performer and for different number
of TCP connections, different policies exhibit improved
performance over other policies. Table I shows
the best performing policies in terms of response time, 
under different number of TCP connections.

\end{document}